\documentclass[aps,pre,twocolumn,showpacs,superscriptaddress]{revtex4}
\usepackage{amssymb}
\usepackage{bbm}
\usepackage{indentfirst}
\usepackage{graphicx}
\usepackage{amsmath,amssymb}
\usepackage{bpchem}

\usepackage{color}

\begin{document}

\title{Double-temperature ratchet model and current reversal of coupled Brownian motors}
\author{Chen-Pu Li}
\affiliation{ Department of Physics, Beijing Normal University, Beijing 100875, China}
\affiliation{ College of Science, Hebei University of Architecture, Zhang Jiakou 075000, China}
\author{Hong-Bin Chen}
\affiliation{Institute of Systems Science (ISS), Huaqiao University, Xiamen 361021, China}
\affiliation{College of Information Science and Engineering, Huaqiao University, Xiamen 361021, China}
\author{Zhi-Gang Zheng}
\email[]{zgzheng@hqu.edu.cn.}
\affiliation{Institute of Systems Science (ISS), Huaqiao University, Xiamen 361021, China}
\affiliation{College of Information Science and Engineering, Huaqiao University, Xiamen 361021, China}
\date{\today}

\begin{abstract}
On the basis of the transport features and experimental phenomena observed in studies of molecular motors, we propose a double-temperature ratchet model of coupled motors to reveal the dynamical mechanism of cooperative transport of motors with two heads, where the interactions and asynchrony between two motor heads are taken into account. We investigate the collective unidirectional transport of coupled system and find that the direction of motion can be reversed under certain conditions. Reverse motion can be achieved by modulating the coupling strength, coupling free length, and asymmetric coefficient of the periodic potential, which is understood in terms of the effective potential theory. The dependence of the directed current on various parameters is studied systematically. Directed transport of coupled Brownian motors can be manipulated and optimized by adjusting the pulsation period or the phase shift of the pulsation temperature.
\end{abstract}

\pacs {05.45.Xt, 05.40.-a, 05.60.-k£Â02.50.Ey}

\maketitle
\section{Introduction}\label{secone}
The directed transport of Brownian motors in periodic structures with the help of fluctuations with zero mean has long been an important problem and has been widely studied~\cite{ref1,ref2,ref3,ref4}. In recent years, much attention has been given to theoretical studies of directed transport of coupled Brownian motors in many different scientific areas, such as molecular motors in biological systems, the divergence of two polymers on a surface, and Josephson junction arrays, to name but a few~\cite{ref5,ref6,ref7,ref8,ref9,ref10,ref11,ref12,ref13,ref14,ref15,ref16}. A nonequilibrium environment and broken symmetry of the system are indispensable conditions for directional transport of particles in a ratchet model. Symmetry breaking generally includes the symmetry breaking of the periodic potential field, the symmetry breaking induced by nonequilibrium perturbations, and the breaking induced by mutual coupling between elements of the system ~\cite{ref17,ref18,ref19,ref20,ref21,ref22,ref23,ref24,ref25,ref26,ref27}.

In this work, we consider two interacting Brownian motors in a spatially periodic potential. We are motivated by experimental observations of the motion patterns of molecular motors (protein motors). It has been found that most protein motors possess a dimer structure in which each motor protein is composed of two interacting identical monomers, and each monomer experiences cyclic ATP hydrolysis~\cite{ref28,ref29}. It has been found experimentally that the coupling between two motor proteins, which do not act independently but alternate in a sequential manner such that their catalytic cycles are out of phase, plays a significant role in achieving directed motion and even reversed motion. We noticed an important fact related to the symmetry breaking of dimer molecular motors; i.e., it was also found that the hydrolysis processes of two heads are continuous while they are  asynchronous~\cite{ref30,ref31}. This gives us a hint for establishing a double-temperature ratchet model of coupled motors in an asymmetric potential field.

In this paper, we analyze the effects of several parameters, such as the coupling strength, asymmetry coefficient of the potential, pulsation period, and phase shift of the temperature, on the average velocity of coupled motors. Furthermore, the dynamical mechanism and reverse behavior of the coupled motor are reasonably explained using effective potential theory in the strong-coupling case.

\section{ The Coupled Double-temperature Brownian Motor Model}\label{sectwo}
\subsection{Dynamical model}\label{subsecone}
We consider the overdamped Brownian motion of two coupled Brownian motors in contact with two reservoirs with different temperatures in asymmetric periodic potentials. The equations of motion of two mutually coupled motors can be written as
\begin{equation}\label{equ:one}
\dot{x_{i}}=-\frac{\partial{V(x_{i})}}{\partial{x_{i}}}-\frac{\partial{U_{0}}}{\partial{x_{i}}}+\xi_{i}(t),\quad i=1,2,
 \end{equation}
where $x_{i}$ is the coordinate of the $i$-th motor, and $V(x)$ is the asymmetric periodic potential originating from the interaction between the motor and the track. The following simplified form of $V(x)$ is chosen:
\begin{equation}\label{equ:2}
V(x)=-V_{0}[\sin{(\frac{2\pi}{L}x)}+\frac{\Delta}{4}\sin{(\frac{4\pi}{L}x)}],
\end{equation}
where $L$ is the spatial period of the potential field $V(x)$, which is shown in Fig.1, and $\Delta$ is the asymmetry coefficient of the potential field $V(x)$. $U_{0}(x_{1},x_{2})$ denotes the interaction potential between the two motors, which is set to have the following simple harmonic form:
\begin{equation}\label{equ:3}
U_{0}(x_{1},x_{2})=\frac{1}{2}k(x_{1}-x_{2}-a)^2,
\end{equation}
where $k$ is the coupling strength, and $a$ is the coupling free length. The influence of the two reservoirs on the Brownian motors is described in terms of the noises $\xi_{1}(t)$ and $\xi_{2}(t)$, which are assumed to be independent, unbiased Gaussian white noises with
\begin{equation}\label{equ:4}
\begin{split}
&\langle\xi_{i}(t)\rangle=0,\\
&\langle\xi_{i}(t)\xi_{j}(t^\prime)\rangle=2k_{B}T_{i}(t)\delta_{ij}\delta(t-t^\prime),\quad i,j=1,2,
\end{split}
\end{equation}
where $k_{B}T$ is the thermal energy, and $T_{1}(t)$, $T_{2}(t)$ are the following modulated harmonically varying functions:
\begin{equation}\label{equ:5}
\begin{split}
&T_{1}(t)=T_{0}[1+A\sin{(\frac{2\pi}{t_{0}}t)}]^{2},\\
&T_{2}(t)=T_{0}[1+A\sin{(\frac{2\pi}{t_{0}}t+\Delta\theta)}]^{2},
\end{split}
\end{equation}
where $t_{0}$ is the pulsation period of the temperature, and $\Delta\theta$ is the phase shift between two temperature fluctuations. The mismatch between the two temperatures denotes biologically the different ATP hydrolysis states of the two motors. Throughout this paper, we set the parameters $T_{0}$ = 0.5 and $A$ = 0.8.
\begin{figure}[h!]
  \centering
  \includegraphics[width=\linewidth]{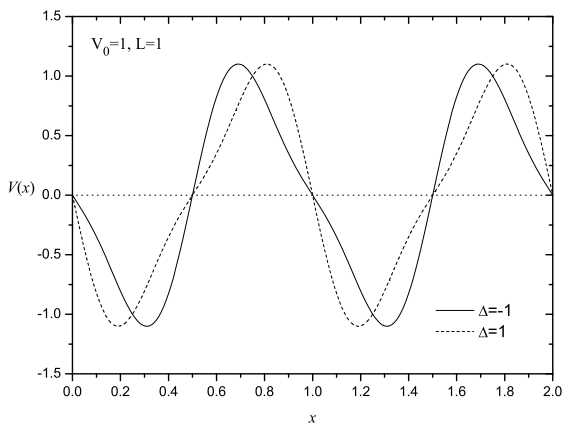}\\
  \caption{ \label{Fig1}the asymmetric periodic potential $V(x)$ which is originated from the interaction between the motor and the track.}
\end{figure}
\subsection{Adiabatic elimination and mass-center dynamics in strong-coupling case}\label{subsectwo}
It is difficult to directly analyze the cooperative ratcheting effect of two coupled Brownian motors theoretically. An effective method for dealing with the cooperative ratcheting effect of coupled motors is to decrease the degrees of freedom of the system to obtain a low-dimensional description. By introducing the mass-center coordinate $X$=$(x_{1}+x_{2})/2$ and the relative coordinate $Y$=$x_{1}-x_{2}$, one can transform Eq. (1) to
\begin{equation}\label{equ:6}
\begin{split}
&\dot{X}=-\frac{1}{2}\frac{\partial [V(X+\frac{Y}{2})+V(X-\frac{Y}{2})]}{\partial X}\\
&-\frac{\partial [V(X+\frac{Y}{2})-V(X-\frac{Y}{2})]}{\partial Y}+\frac{1}{2}(\xi_{1}(t)+\xi_{2}(t)),
\end{split}
\end{equation}

\begin{equation}\label{equ:7}
\begin{split}
&\dot{Y}=-\frac{\partial [V(X+\frac{Y}{2})-V(X-\frac{Y}{2})]}{\partial X}\\
&-2\frac{\partial [V(X+\frac{Y}{2})+V(X-\frac{Y}{2})]}{\partial Y}-2k(Y-a)+\xi_{1}(t)-\xi_{2}(t).
\end{split}
\end{equation}
To understand the effect of the coupling between two Brownian motors on the directed transport of the system, it is instructive to study the dynamics in the limit of a large but finite stiffness $k$. In this situation, the term $-2k(Y-a)$ in Eq. (7) implies a much faster decay of the coordinate $Y$ compared with the relaxation of the variable $X$. This analysis indicates that $Y$ is a fast variable and can be adiabatically eliminated in terms of the slaving principle proposed by Haken~\cite{ref10,ref27,ref32}. Therefore, in $k\rightarrow\infty$, the dynamical equation of the mass center of coupled motors can be described as
\begin{equation}\label{equ:8}
\begin{split}
&\dot{X}=-\frac{1}{2}\frac{\partial [V(X+\frac{a}{2})+V(X-\frac{a}{2})]}{\partial X}\\
&+\frac{1}{2}(\xi_{1}(t)+\xi_{2}(t)),
\end{split}
\end{equation}
with $Y\approx a$.

The dynamical equation (8) can be simply expressed as
\begin{equation}\label{equ:9}
\dot{X}=f(X)+q(t),
\end{equation}
with
\begin{eqnarray*}
f(X) &=& -\frac{1}{2}\frac{\partial [V(X+\frac{a}{2})+V(X-\frac{a}{2})]}{\partial X}, \\
q(t) &=& \frac{1}{2}(\xi_{1}(t)+\xi_{2}(t)).
\end{eqnarray*}
Theoretically, this enables one to qualitatively calculate the effect of multiple parameters, such as the coupling free length $a$, the modulation period $t_{0}$ of the two temperatures, and the phase shift $\Delta\theta$ between the two temperatures, on the ratchet motion of the coupled Brownian motors in terms of the dynamics of a single Brownian motor (the mass-center dynamics).

According to the analyses in reference [4] of the temperature ratchet of a single Brownian motor, when the period $t_{0}$ of the temperature fluctuation tends to infinity, the temperature can be regarded as an approximate constant in a small time interval, where the average velocity of a single Brownian motor is zero in the periodic potential with Gaussian white noise. When the period $t_{0}\ll1$, one can readily find that
\begin{equation}\label{equ:10}
\langle\dot{X}\rangle=t_{0}^{2}B_{k}\int_{0}^{L}dX(V_{k}^{'}(X)[V_{k}^{''}(X)]^{2})+o(t^{3}),
\end{equation}
with
\begin{eqnarray*}
B_{k}=\frac{4L\int_{0}^{1}dh[\int_{0}^{h}d\hat{h}(\frac{1-\hat{T}(\hat{h})}{\bar{T}})]}{\eta^{3}\int_{0}^{L}dX(e^{\frac{V_{k}(X)}{k_{B}\bar{T}}})\int_{0}^{L}dX(e^{-\frac{V_{k}(X)}{k_{B}\bar{T}}})}
\end{eqnarray*}
and
\begin{eqnarray*}
&&\bar{T}=\frac{1}{t_{0}}\int_{0}^{t_{0}}dtT(t)=\int_{0}^{1}dh\hat{T}(h),\\
&&\hat{T}(h)=T(t)=T(t_{0}h).
\end{eqnarray*}
\subsection{Effective potential theory in strong-coupling case}\label{subsecthree}
To explain the current reversal of the ratchet system, it is convenient to introduce the effective potential~\cite{ref10,ref24,ref27}. In the above discussion, we know that in the strong-coupling situation, the dynamics of the relative coordinate $Y$ occurs on a much faster time scale than that of the mass-center coordinate $X$. One can obtain the effective potential $V_{eff} (X)$ of $X$ by replacing the $X$- and $Y$-dependent potential with the potential averaged with respect to the fast relative coordinate $Y$ as
\begin{equation}\label{equ:11}
V_{eff}(X)=-\frac{1}{2}k_{B}\bar{T}\ln(\int_{-\infty}^{\infty}dY\rho(X,Y)),
\end{equation}
and
\begin{equation}\label{equ:12}
\rho(X,Y)=e^{-\frac{U(X,Y)}{k_{B}\bar{T}}},
\end{equation}
\begin{equation}\label{equ:13}
U(X,Y)=\frac{1}{2}k(Y-a)^{2}+V(X+\frac{Y}{2})+V(X-\frac{Y}{2}).
\end{equation}

In the following discussions, we continue to study the collective directed transport of coupled Brownian motors using numerical simulations and make comparisons with theoretical discussions. In numerical simulations of the stochastic dynamics, the second-order Runge--Kutta numerical simulation algorithm is adopted~\cite{ref33,ref34}; the number of ensembles $N$ is 1000, and the time step $dt$ is 0.001. Throughout the numerical simulations, we set the parameter $V_{0}$ to 1 and $L$ to 1. The average velocity or the current can be obtained by obtaining both ensemble and time averages of the instantaneous velocity as
\begin{equation}\label{equ:14}
<v_{c}>=<\dot{X_{c}}>=\lim_{t\rightarrow\infty}\frac{1}{Nt_{0}}\sum_{i=1}^{N}\int_{0}^{t_{0}}dt^{'}\dot{X}_{ci}(t^{'}).
\end{equation}
\section{ Collective Directed Transport and Current Reversal}\label{secthree}
In this section, we analyze the influence of the coupling strength $k$ of the coupled Brownian motors, the coupling free length $a$, and the phase shift $\Delta\theta$ of the two motor heads on the average velocity. Theoretically, one can discuss the current reversal using the effective potential theory for the strong-coupling case.

\subsection{Current reversal induced by the coupling strength}\label{subsecone}
The coupling strength $k$ plays a significant role in directed transport of the ratchet system. Fig.2(a) shows three curves of the average velocity against the coupling strength $k$ with phase shifts $\Delta\theta=\pi$, $\pi/2$, and 0. Fig.2(b) presents the curves of the effective potential with coupling strengths $k$=0, 300, and 1000 (corresponding to the strong-coupling limit $k\rightarrow\infty$ in the theoretical analysis). For every curve in Fig.2(a), the sign of the velocity, which corresponds to the direction of motion of the coupled motors, can be reversed when the coupling strength $k$ reaches a certain value.

For weak coupling [e.g., $0.01<k<1$ in Fig.2(a)], the average velocity $v\approx-0.22$. This is because in the weak-coupling case, the motion can be considered as a simple combination of that of two single motors, where each particle is immersed in a common periodic potential $V(x)$ and the Gaussian white noise, and the effective potential $V_{eff}(X)$ [e.g., the curve with $k$ = 0 in Fig.2(b)] of the mass center could be considered as the potential $V(x)$ of a single motor; the motion of two weakly coupled motors is consistent with the relatively large gradient of the potential $V(x)$ in the temperature ratchet model of a single motor~\cite{ref10}. With increasing coupling strength $k$ [e.g., $k = 1-40$ in Fig.2(a)], the negative mean velocity gradually increases and tends to zero, which shows that the coupling between the two coupled motors affects the symmetry breaking of the entire system. When the coupling is large enough [e.g., $k=40-1000$ in Fig.2(a)], the symmetry breaking is opposite to that when the coupling is weak [$k\sim$ 0 to 40 in Fig.2(a)], such that the effective potential $V_{eff}(X)$ in the strong-coupling case exhibits the opposite tendency of that in the weak-coupling case in Fig.2(b). Finally, the positive mean velocity reaches saturation with increasing coupling strength $k$.

In addition, when $k<40$, the three curves in Fig.2(a) almost overlap, and the effect of the phase shift $\Delta\theta$ on the average velocity is very small. That is because of the single-particle behavior and the fact that the mean velocity is independent of $\Delta\theta$ for the weak-coupling case. When $k>40$, the mean velocity decreases with increasing $\Delta\theta$ in the range of [0,$\pi$] and tends to zero for $\Delta\theta=\pi$. This can be interpreted in terms of the formulas for the two temporally modulated temperatures $T_{1}(t)$=$T_{0}[1+A\sin(2\pi t/t_{0})]^{2}$ and $T_{2}(t)$=$T_{0}[1+A\sin(2\pi t/t_{0}+\Delta\theta)]^{2}$. In the temperature ratchet model, the maximum temperature represents the minimum binding of the potential well to the coupled motors, and the minimum temperature corresponds to the strongest binding. According to the two formulas with $\Delta\theta=\pi$, one temperature of the coupled motors reaches the maximum value when the other one is at the minimum, indicating that the coupling reduces the motion of the coupled motor in this case. For $\Delta\theta=0$, the temperature fluctuation of both motors is synchronized, which means that the average velocity reaches the maximum value because the coupling could enhance the motion.
\begin{figure}[h!]
  \centering
  \includegraphics[width=\linewidth]{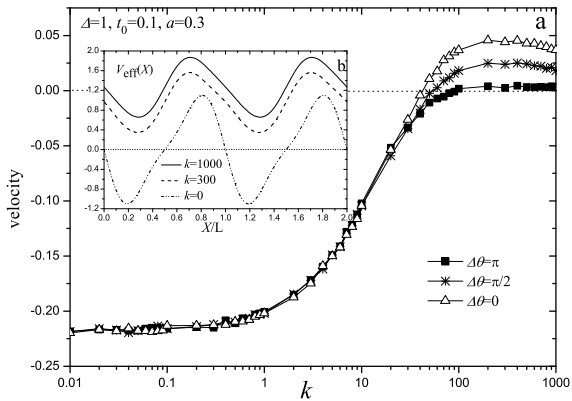}\\
  \caption{ \label{Fig2}the average velocity as a function of the coupling strength $k$.}
\end{figure}

\subsection{Current reversal induced by changing the coupling free length}\label{subsectwo}
The presence of coupling between two motor heads not only provides cooperative directed motion, but also significantly affects the current. The direction of the current can even be determined by the free length of the coupled motors, where the current reversal can be interpreted well in terms of the effective potential theory.

In Fig.3(a), we give the average velocity versus the coupling free length $a$ with the other parameters given in the plot, where the four curves correspond to different coupling strengths $k$ and phase shifts $\Delta\theta$ between the two temperatures, i.e., ($k$, $\Delta\theta$)=(1, 0), (300, 0), (300, $\pi/2$), and (300, $\pi$). It can be found from Fig.3(a) that when the coupling strength $k$ is small (e.g., $k=1$), the coupling free length $a$ cannot greatly influence the average velocity, which has an almost constant negative value (e.g., $v= -0.2$) for all values of $a$. This is because the motion of coupled motors can be considered as a simple combination of that of two single motors in weak-coupling cases. For the three curves with strong coupling strengths $k$ (e.g., $k=300$) between the two motors in Fig.3(a), the average velocity first increases and then decreases with $a$ in the range of 0 to $L/2$, and the velocity is reversed when $a=L/4$. In addition, these curves are symmetric about $a=L/2$. In Fig.3(b), the average velocity versus the coupling free length $a$ is theoretically computed according to Eq. (10) with the parameters given in the plot. Although this result differs from the simulation results in Fig.3(a), the theoretical results can still qualitatively display the effects of the coupling free length $a$ on the average velocity. Note that the theoretical results agree well with that of the simulation only when $t_{0}$ is close to zero.
\begin{figure}[h!]
  \centering
  \includegraphics[width=\linewidth]{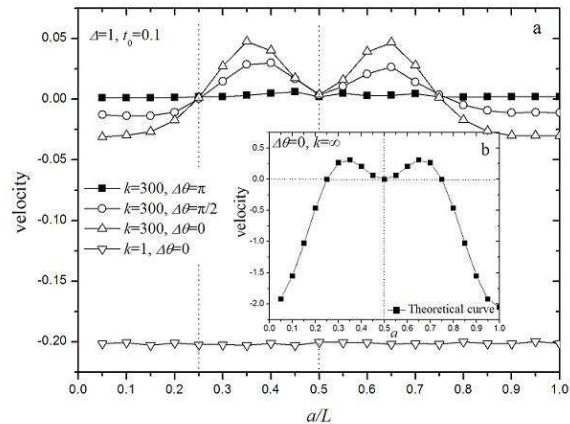}\\
  \caption{ \label{Fig3}the average velocity as function of the coupling free length $a$.}
\end{figure}

In Fig.4(a), (b), (c), and (d), we give the effective potential $V_{eff}(X)$ for coupling free lengths $a$ of 0.1, 0.25, 0.3, and 0.5, respectively, where the other parameters are $k$ = 300, $\Delta$ = 1, $t_{0}$ = 0.1, and $\Delta\theta$=0. We introduce an effective asymmetric coefficient $\Delta_{eff}$ of the effective potential $V_{eff}(X)$, where the effective asymmetric coefficient $\Delta_{eff}=(L2-L1)/L$, which is shown in Fig.4(a). In the strong-coupling case, the asymmetric coefficient of the effective potential $V_{eff}(X)$ periodically changes against the coupling free length $a$ with $L$ in one cycle in Fig.4(a), (b), (c), and (d). Fig.4 shows that the asymmetric coefficient $\Delta_{eff}$ is less than zero when $0<a<L/4$, is greater than zero when $L/4<a<L/2$, and equals zero when $a=L/4$ and $L/2$. The cyclic behavior of the average velocity agrees well with the change in the asymmetric coefficient of the effective potential $V_{eff}(X)$ for the strong-coupling situations shown in Fig.4, which naturally leads to the periodic variation of the average velocity.

The symmetry of the average velocity about $a=L/2$ can be understood well by analyzing the space-time transformation invariance of the dynamical Eq. (1)~\cite{ref25,ref27,ref35,ref36}. Rewriting Eq. (1) with a superscript corresponding to the value of the coupling free length $a$ yields

\begin{figure}[h!]
  \centering
  \includegraphics[width=\linewidth]{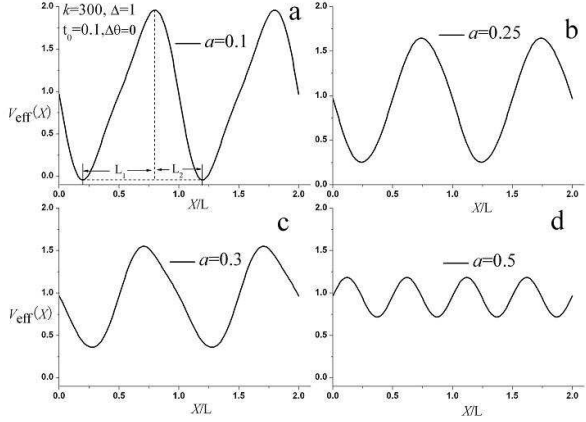}\\
  \caption{ \label{Fig4}the curves of effective potential with different the coupling free length $a$.}
\end{figure}

\begin{equation}\label{equ:15}
\begin{split}
&\dot{x}_{1}^{(a)}=-\frac{\partial V(x_{1})}{\partial x_{1}}-k(x_{1}-x_{2}-a)+\xi_{1}(t),\\
&\dot{x}_{2}^{(a)}=-\frac{\partial V(x_{2})}{\partial x_{2}}+k(x_{1}-x_{2}-a)+\xi_{2}(t).
\end{split}
\end{equation}
By replacing the free length $a$ with $L-a$, we obtain
\begin{equation}\label{equ:16}
\begin{split}
&\dot{x}_{1}^{(L-a)}=-\frac{\partial V(x_{1})}{\partial x_{1}}-k(x_{1}-x_{2}-L+a)+\xi_{1}(t),\\
&\dot{x}_{2}^{(L-a)}=-\frac{\partial V(x_{2})}{\partial x_{2}}+k(x_{1}-x_{2}-L+a)+\xi_{2}(t).
\end{split}
\end{equation}
By inserting $x_{11}=x_{1}-L$ into Eq. (16), we have
\begin{equation}\label{equ:17}
\begin{split}
&\dot{x}_{11}^{(L-a)}=-\frac{\partial V(x_{11})}{\partial x_{11}}-k(x_{11}-x_{2}+a)+\xi_{1}(t),\\
&\dot{x}_{2}^{(L-a)}=-\frac{\partial V(x_{2})}{\partial x_{2}}+k(x_{11}-x_{2}+a)+\xi_{2}(t).
\end{split}
\end{equation}
By further transforming Eq.(17) by first $x_{11} \rightarrow x_{1}$ and then $x_{1}\leftrightarrow x_{2}$, we obtain
\begin{equation}\label{equ:18}
\begin{split}
&\dot{x}_{2}^{(L-a)}=-\frac{\partial V(x_{2})}{\partial x_{2}}+k(x_{1}-x_{2}-a)+\xi_{1}(t),\\
&\dot{x}_{1}^{(L-a)}=-\frac{\partial V(x_{1})}{\partial x_{1}}-k(x_{1}-x_{2}-a)+\xi_{2}(t).
\end{split}
\end{equation}
Taking into account that $\xi_{1}$ and $\xi_{2}$ have the same statistical properties, we finally obtain that
\begin{equation}\label{equ:19}
\begin{split}
&\dot{x}_{2}^{(L-a)}=-\frac{\partial V(x_{2})}{\partial x_{2}}+k(x_{1}-x_{2}-a)+\xi_{2}(t)\\
&=\dot{x}_{2}^{(a)},\\
&\dot{x}_{1}^{(L-a)}=-\frac{\partial V(x_{1})}{\partial x_{1}}-k(x_{1}-x_{2}-a)+\xi_{1}(t)\\
&=\dot{x}_{1}^{(a)}.
\end{split}
\end{equation}
That is,
\begin{equation}\label{equ:20}
\begin{split}
&v^{(L-a)}=(\dot{x}_{1}^{(L-a)}+\dot{x}_{2}^{(L-a)})/2=v^{(a)}\\
&=(\dot{x}_{1}^{(a)}+\dot{x}_{2}^{(a)})/2.
\end{split}
\end{equation}
This result makes it possible to interpret the symmetry of the average velocity against the coupling free length $a$, as shown in Fig.3.

\subsection{Current reversal induced by the potential asymmetry coefficient}\label{subsecthree}
In studies of Brownian ratchets, the asymmetry coefficient $\Delta$ of the potential $V(x)$ significantly affects the directional current. In the above discussions, we observed reversal of the directed motion by modulating the coupling strength $k$ and coupling free length $a$. In fact, the direction can also be determined by the asymmetric coefficient $\Delta$ for a given $a$ and $k$. In Fig.5(a), the average velocity against the asymmetric coefficient $\Delta$ is plotted for ($k$, $\Delta\theta$) = (1, 0), (300, 0), and (300, $\pi$), where the other parameters are $t_{0}$ = 0.1 and $a$ = 0.3.
\begin{figure}[h!]
  \centering
  \includegraphics[width=\linewidth]{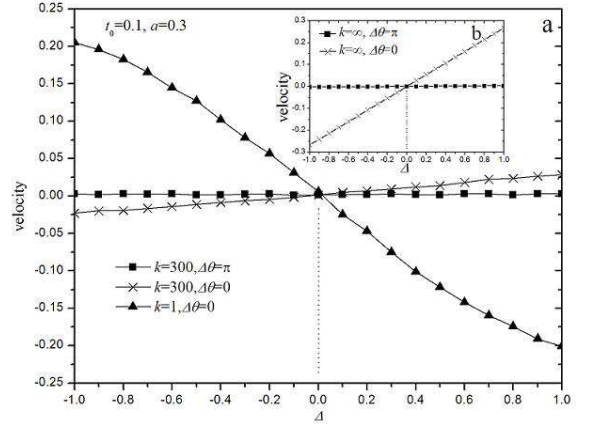}\\
  \caption{ \label{Fig5}the average velocity varying with the asymmetric coefficient $\Delta$ of $V(x)$.}
\end{figure}

In the weak-coupling case (e.g., $k=1$), the curve for $k=1$, $\Delta\theta=0$ in Fig.5(a) shows that $v>0$ when $\Delta<0$, $v<0$ when $\Delta>0$, and $v=0$ when $\Delta=0$. Thus, the size and direction of the average velocity are determined by the asymmetric coefficient $\Delta$ in this case. In addition, the average velocity decreases approximately linearly with increasing asymmetric coefficient $\Delta$. According to the above analysis, the motion of coupled Brownian motors can be approximately considered as the motion of a single Brownian motor immersed in the periodic potential $V(x)$ and Gaussian white noise when the coupling strength $k$ is small, where the direction of the velocity is determined by $\Delta$ in the temperature ratchet model of a single particle. In the strong-coupling case (e.g., $k=300$), the average velocity, the size and direction of which are also determined by the asymmetric coefficient $\Delta$, increases linearly with increasing asymmetric coefficient $\Delta$. The curves of the average velocity are reversed compared to those of the weak-coupling case (e.g., $k=1$). Moreover, the average velocity for $\Delta\theta=0$ is larger than that for $\Delta\theta=\pi$ when $k=300$. Similarly, Fig.5(b) gives the theoretical curve of the average velocity versus the asymmetric coefficient $\Delta$ according to Eq. (10), which agrees qualitatively with the simulation results shown in Fig.5(a).
\begin{figure}[h!]
  \centering
  \includegraphics[width=\linewidth]{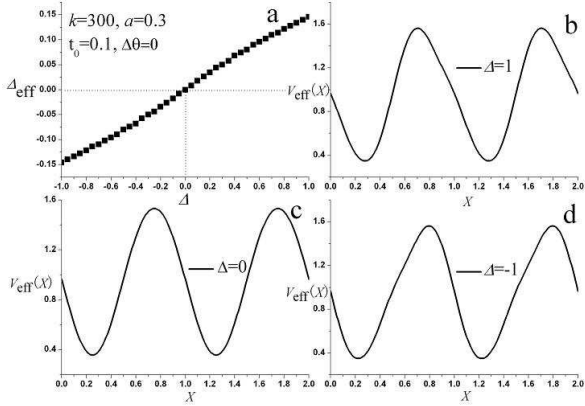}\\
  \caption{ \label{Fig6}the effective potential $V_{eff}(X)$ with different asymmetric coefficient $\Delta$.}
\end{figure}

The effects of the asymmetric coefficient $\Delta$ on the average velocity can be explained by the effective potential theory for the strong-coupling case. When $\Delta_{eff}>0$, the average velocity $v>0$, and when $\Delta_{eff}<0$, the velocity $v<0$ according to the dynamical mechanism of the temperature ratchet model. The curve of the effective asymmetric coefficient $\Delta_{eff}$ versus the asymmetric coefficient $\Delta$ is plotted in Fig.6(a). One can easily find that the effective asymmetric coefficient increases linearly with the asymmetric coefficient $\Delta$, and when $\Delta=0$, the effective asymmetric coefficient $\Delta_{eff}$ is also zero. Fig.6(b), (c), and (d) show the effective potential for different asymmetric coefficients $\Delta$ given in the plot. Fig.6 shows that the asymmetry of $V_{eff}(X)$ when $\Delta<0$ is opposite to that when $\Delta>0$, and the effective potential $V_{eff}(X)$ is symmetric when $\Delta=0$. The asymmetry of the effective potential decreases and it is close to symmetric when $\Delta$ approaches zero. These results are in agreement with the dependence of the averaged velocity on the asymmetric coefficient $\Delta$ for the strong-coupling case shown in Fig.5(a).

Moreover, it is noteworthy that every curve in Fig.5(a) is antisymmetric about $\Delta=0$ for different coupling cases. This can be explained well through an analysis of the space-time transformation invariance. As we did above, we rewrite Eq. (1), this time with a superscript corresponding to the value of the asymmetric coefficient $\Delta$, and obtain
\begin{equation}\label{equ:21}
\begin{split}
&\dot{x}_{1}^{(\Delta)}=-\frac{\partial V(x_{1},\Delta)}{\partial x_{1}}-k(x_{1}-x_{2}-a)+\xi_{1}(t),\\
&\dot{x}_{2}^{(\Delta)}=-\frac{\partial V(x_{2},\Delta)}{\partial x_{2}}+k(x_{1}-x_{2}-a)+\xi_{2}(t).
\end{split}
\end{equation}
Adding the two formulas in Eq. (21), one naturally gets
\begin{equation}\label{equ:22}
\begin{split}
&\dot{x}_{1}^{(\Delta)}+\dot{x}_{2}^{(\Delta)}=-[\frac{\partial V(x_{1},\Delta)}{\partial x_{1}}+\frac{\partial V(x_{2},\Delta)}
{\partial x_{2}}]\\
&+\xi_{1}(t)+\xi_{2}(t).
\end{split}
\end{equation}
By replacing the free length $\Delta$ with $L-\Delta$, we have
\begin{equation}\label{equ:23}
\begin{split}
&\dot{x}_{1}^{(-\Delta)}+\dot{x}_{2}^{(-\Delta)}=-[\frac{\partial V(x_{1},-\Delta)}{\partial x_{1}}+\frac{\partial V(x_{2},-\Delta)}
{\partial x_{2}}]\\
&+\xi_{1}(t)+\xi_{2}(t).
\end{split}
\end{equation}
Then, by putting $ V(x)=-V_{0}[\sin(2\pi x/L)+(\Delta/4)\sin(4\pi x/L)]$ into Eq. (23), we obtain
\begin{equation}\label{equ:24}
\begin{split}
&\dot{x}_{1}^{(-\Delta)}+\dot{x}_{2}^{(-\Delta)}=\frac{2\pi V_{0}}{L}[\cos(\frac{2\pi x_{1}}{L})-\frac{\Delta}{2}\cos(\frac{4\pi x_{1}}{L})\\
&+\cos(\frac{2\pi x_{2}}{L})-\frac{\Delta}{2}\cos(\frac{4\pi x_{2}}{L})]+\xi_{1}(t)+\xi_{2}(t),
\end{split}
\end{equation}
\begin{equation}\label{equ:25}
\begin{split}
&\dot{x}_{1}^{(-\Delta)}+\dot{x}_{2}^{(-\Delta)}=-\frac{2\pi V_{0}}{L}[\cos(\frac{2\pi x_{1}}{L}-\pi)+\frac{\Delta}{2}\cos(\frac{4\pi x_{1}}{L})\\
&+\cos(\frac{2\pi x_{2}}{L}-\pi)+\frac{\Delta}{2}\cos(\frac{4\pi x_{2}}{L})]+\xi_{1}(t)+\xi_{2}(t).
\end{split}
\end{equation}
By further inserting $x_{1}=x_{11}+L/2$ and $x_{2}=x_{21}+L/2$ into Eq. (25), we have
\begin{equation}\label{equ:26}
\begin{split}
&\dot{x}_{1}^{(-\Delta)}+\dot{x}_{2}^{(-\Delta)}=-\frac{2\pi V_{0}}{L}[\cos(\frac{2\pi x_{11}}{L})+\frac{\Delta}{2}\cos(\frac{4\pi x_{11}}{L})\\
&+\cos(\frac{2\pi x_{21}}{L})+\frac{\Delta}{2}\cos(\frac{4\pi x_{21}}{L})]+\xi_{1}(t)+\xi_{2}(t).
\end{split}
\end{equation}
By then replacing $x_{11}\rightarrow x_{1}$ and $x_{21}\rightarrow x_{2}$, we find that
\begin{equation}\label{equ:27}
\begin{split}
&\dot{x}_{1}^{(-\Delta)}+\dot{x}_{2}^{(-\Delta)}=-\frac{2\pi V_{0}}{L}[\cos(\frac{2\pi x_{1}}{L})+\frac{\Delta}{2}\cos(\frac{4\pi x_{1}}{L})\\
&+\cos(\frac{2\pi x_{2}}{L})+\frac{\Delta}{2}\cos(\frac{4\pi x_{2}}{L})]+\xi_{1}(t)+\xi_{2}(t)\\
&=[\frac{\partial V(x_{1},\Delta)}{\partial x_{1}}+\frac{\partial V(x_{2},\Delta)}
{\partial x_{2}}]+\xi_{1}(t)+\xi_{2}(t)\\
&=-(\dot{x}_{1}^{(\Delta)}+\dot{x}_{2}^{(\Delta)}).
\end{split}
\end{equation}
Taking into account that the Gaussian white noises $\xi_{1}$ and $\xi_{2}$ have the same statistical properties, and comparing with Eq. (22), eventually we find that
\begin{equation}\label{equ:28}
\begin{split}
&v^{(-\Delta)}=(\dot{x}_{1}^{(-\Delta)}+\dot{x}_{2}^{(-\Delta)})/2=-(\dot{x}_{1}^{(\Delta)}+\dot{x}_{2}^{(\Delta)})/2\\
&=-v^{(\Delta)}.
\end{split}
\end{equation}
This implies that the relationship between the velocity of the coupled Brownian motors and the asymmetric coefficient $\Delta$ is antisymmetric. For the special value $\Delta=0$, this leads to unbiased motion of the coupled motors, and one has $v^{(0)}=0$. These results agree well with the effective potential analysis.
\section{ Optimization and Manipulation of Collective Directed Transport}\label{secfour}
It is interesting to investigate the dependence of the directed transport of the system on the pulsation period $t_{0}$ and the phase shift $\Delta\theta$ between the two motor heads. In this section, we systematically determine how to optimize the motor motion by modulating the pulsation period and phase shift.
\subsection{Effect of the pulsation period of the temperature}\label{subsecone}
The average velocity of the coupled motors against $t_{0}$ is plotted in Fig.7 for different coupling strengths $k$ and phase shifts $\Delta\theta$, with $\Delta=1$ and $a=0.3$. The averaged velocity is negative when $k$ is small (e.g., $k=1$), whereas the direction of motion is reversed when the coupling strength $k$ is large (e.g., $k=300$). These results are consistent with the discussion above in terms of the effective potential analysis shown in Fig.2(b). For the two curves with $k=300$, the average velocity is larger for the phase shift $\Delta\theta=0$ than for $\Delta\theta=\pi$, and the velocity is nearly zero for the latter.
\begin{figure}[h!]
  \centering
  \includegraphics[width=\linewidth]{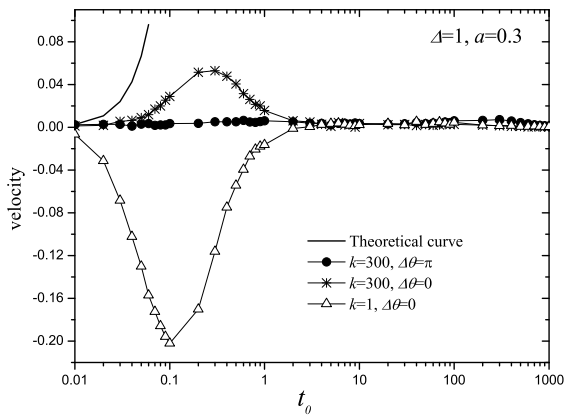}\\
  \caption{ \label{Fig7}the average velocity as the function of the pulsating period $t_{0}$ of the temperature $T$.}
\end{figure}

In Fig.7, we plot the average velocity as a function of the pulsation period $t_{0}$ of the temperature $T$. All the curves of the mean velocity clearly have an optimal pulsation period $t_{max}$ having the largest velocity, and the averaged velocity approaches zero when $t_{0}$ tends to zero and infinity. This interesting behavior can be interpreted as follows. For an infinitely large pulsation period $t_{0}$, the temperature pulsates so slowly that it can be regarded as a constant. In this case, the coupled Brownian motors are immersed in a stationary periodic potential and a white Gaussian noise with a constant intensity, where the system has a null directed current regardless of the coupling strength $k$. In contrast, when $t_{0}$ tends to zero, the temperature fluctuates too rapidly, and the change in the configuration of the coupled Brownian motors always lags behind the rapid temperature fluctuation. This also leads to an absence of directed motion. Moreover, the solid curve in Fig.7 gives the theoretical result according to formula (10) with $t_{0}$ and $k$ tending to zero and infinity. The theoretical results agree well with that of the simulation when $t_{0}$ is close to zero.
\subsection{Effect of the phase shift between the two temperatures}\label{subsectwo}
The phase shift $\Delta\theta$ between the temperatures of the two motors significantly affects the current. In Fig.8(a), the average velocity versus the phase shift $\Delta\theta$ is plotted for coupling strengths of $k$ = 1 and 300, and the other parameters are $\Delta$ = 1, $a$ = 0.3, and $t_{0}$ = 0.1. For a very small coupling strength $k$ (e.g., $k$ = 1), the effect of the phase shift $\Delta\theta$ on the average velocity is almost negligible, and the velocity is approximately a negative constant, $v=-0.2$. For strong coupling (e.g., $k$ = 300), the phase shift $\Delta\theta$ has important effects on the average velocity, as clearly shown in Fig.8 for the curve with $k$ = 300. Moreover, the average velocity versus the phase shift $\Delta\theta$ is symmetric about $\Delta\theta=\pi$ owing to the temporal periodicity of the temperature $T(t)$ about $2\pi$, where $v$ first decreases and then increase as $\Delta\theta$ increases in the range of 0 to $2\pi$. Fig.8(b) shows the theoretical average velocity versus the phase shift $\Delta\theta$ according to Eq. (10) with the parameters given in the figure, which agrees qualitatively with the simulated data shown in Fig.8(a).
\begin{figure}[h!]
  \centering
  \includegraphics[width=\linewidth]{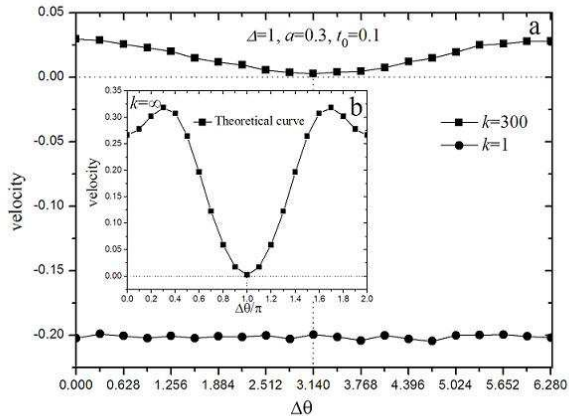}\\
  \caption{ \label{Fig8}The average velocity versus the phase shift $\Delta\theta$ with different the coupling strengths $k$.}
\end{figure}

\section{Concluding Remarks}\label{secfive}
In this paper, we studied the mechanism of collaborative directed transport of elastically coupled Brownian motors in an asymmetric periodic potential under the modulation of two reservoirs with different and asynchronous temperatures. We applied both invariance analysis of the space-time transformation of the coupled dynamical equations and the effective potential theory to study the coupling-induced symmetry breaking and the consequent collective directed transport and current reversal behavior. The dynamical analysis indicated that $\Delta\neq0$ is the precondition of directed motion of the coupled system in this dynamical model, and the presence of coupling between the two motors influences the symmetry breaking of the ratchet system. In the strong-coupling case, the directional transport of the coupled motors can be reversed by modulating the coupling strength, coupling free length, or asymmetry coefficient, which can be effectively illustrated using the effective potential theory and invariance analysis of the space-time transformation. Moreover, the relationships between the average velocity and various parameters such as the pulsation period and the phase shift between the temperatures of the two reservoirs are investigated systematically, and the results provide a valuable way of optimizing and manipulating the collective directed transport by adjusting different parameters in practice.

This work is partially supported by the National Natural Science Foundation of China (Grant Nos.11075016 and 11475022),the Scientific Research project of Zhangjiakou city (Grant Nos.1611064B) and the Scientific Research Funds of Huaqiao University.


\begin{thebibliography}{99}
\bibitem {ref1}P. Reimann, M. Evstigneev, Europhys. Lett.{\bf 78}, 50004 (2007).
\bibitem {ref2}F. Marchesoni, Phys. Rev. E {\bf 56}, 2497 (1997).
\bibitem {ref3}J. D. Bao, Y. Z. Zhuo, Chin. Sci. Bull. {\bf 43}, 1497 (1998).
\bibitem {ref4}P. Reimann, Phys. Rep. {\bf 361}, 57 (2002).
\bibitem {ref5}O. M. Braun, R. Ferrando, and G. E. Tommei, Phys. Rev. E {\bf 68}, 051101 (2003).
\bibitem {ref6}S. Goncalves, C. Fusco, A. R. Bishop, and V. M. Kenkre, Phys. Rev. B {\bf 72}, 195418 (2005).
\bibitem {ref7}E. Heinsalu, M. Patriarca, and F. Marchesoni, Phys. Rev. E {\bf 77}, 021129 (2008).
\bibitem {ref8}A. E. Filippov, J. Klafter, and M. Urbakh, Phys. Rev. Lett. {\bf 92}, 135503 (2004).
\bibitem {ref9}S. Maier, Y. Sang, T. Filleter, M. Grant, R. Bennewitz, E. Gnecco, and E. Meyer, Phys. Rev. B {\bf 72}, 245418 (2005).
\bibitem {ref10}H. Y. Wang and J.D. Bao, Physica A {\bf 374}, 33 (2007).
\bibitem {ref11}J. L. Mateos, Physica A {\bf 351}, 79 (2005).
\bibitem {ref12}S. E. Mangioni and H. S. Wio, Eur. Phys. J. B {\bf 61}, 67 (2008).
\bibitem {ref13}E. M. Craig, M. J. Zuckermann, and H. Linke, Phys. Rev. E  {\bf 73}, 051106 (2006).
\bibitem {ref14}J. Menche and L. Schimansky-Geier, Phys. Lett. A  {\bf 359}, 90 (2006).
\bibitem {ref15}M. Evstigneev, S. V. Gehlen, and P. Reimann, Phys. Rev. E {\bf 79}, 011116(2009).
\bibitem {ref16}C. Lutz, M. Reichert, H. Stark, and C. Bechinger, Europhys.Lett. {\bf 74}, 719 (2006).
\bibitem {ref17}T. F. Gao, B. Q. Ai, Z. G. Zheng, and J. C. Chen, Jour. Stat. Mech. {\bf 09}, 093204 (2016).
\bibitem {ref18}H. Y. Wang, J. D. Bao, Physica A {\bf 389}, 433 (2010).
\bibitem {ref19}Z. G. Zheng, Commun. Theor. Phys. {\bf 43}, 1072 (2005).
\bibitem {ref20} B. O. Yan, R. M. Miura, Y. D. Chen, J. Theor. Bio.{\bf 210}, 141(2001).
\bibitem {ref21}A. Pototsky, N.B. Janson, F. Marchesoni, and S. Savelev, Europhys. Lett. {\bf 88}, 30003 (2009).
\bibitem {ref22}Z. G. Zheng, G. Hu, B. Hu, Phys. Rev. Lett. {\bf 86}, 2273 (2001).
\bibitem {ref23}S.V. Gehlen, M. Evsstigneev, and P. Reimann, Phys.Rev.E {\bf 79}, 031114 (2009).
\bibitem {ref24}H. Y. Wang, J. D. Bao, Physica A {\bf 337}, 13 (2004).
\bibitem {ref25}Z. G. Zheng, M. C. Cross, G, Hu, Phys. Rev. Lett. {\bf 89}, 157102 (2002).
\bibitem {ref26}Z. G. Zheng, H. B. Chen, Europhys. Lett. {\bf 92}, 3004 (2010).
\bibitem {ref27}S.V. Gehlen, M. Evsstigneev, and P. Reimann, Phys. Rev. E {\bf 77}, 031136 (2008).
\bibitem {ref28}A. D. Rogat, K. G. Miler, J. Cell Sci. {\bf 115}, 4855 (2002).
\bibitem {ref29}H. Park, A. Li, L. Q. Chen, A. Houdusse, P. R. Selvin, H. L. Sweeney,  Proc. Natl. Acad. Sci. {\bf 104}, 778 (2007).
\bibitem {ref30}E. M. Delacruz, E. M. Ostap, H. L. Sweeney, J. Biochem. {\bf 276}, 32373 (2001).
\bibitem {ref31}S. Nishikawa, K. Homma, Y. Komori, M.Iwaki, T. Wazawa, A.H. Iwone, J.Saito, R, Ikebe, E. Katayama, T.Yanagida, M.Ikebe, Biochem. Biophs. Res. Commun. {\bf 290}, 311 (2002).
\bibitem {ref32}A. Wunderlin, H. Haken. Zeitschrift fur Physik B Condensed Matter {\bf 44}, 135 (1981).
\bibitem {ref33}J. C. Chen, G. Z. Su, {\it Thermodynamics and statistical physics (Vol.1)} (Science Press, Beijing, 2010)(in Chinese).
\bibitem {ref34}J. D. Bao, {\it Stochastic simulation method of classical and quantum dissipative systems} (Science Press, Beijing, 2009)(in Chinese).
\bibitem {ref35} Z. G. Zheng, {\it Collective behaviors and spatiotemporal dynamics in coupled nonlinear system} (Higher Education Press, Beijing, 2004)(in Chinese).
\bibitem {ref36}H. B. Chen, Q.W. Wang, Z. G. Zheng, Phys. Rev. E {\bf 71}, 031102 (2005).

\end{thebibliography}
\end{document}